**Simultaneous achievement of large anomalous Nernst effect and reduced thermal conductivity in sintered polycrystalline topological Heusler ferromagnets**

Koichi Oyanagi[a,b,c,*], Hossein Sepehri-Amin[b,*], Kenta Takamori[a], Terumasa Tadano[b], Takumi Imamura[a,b,d], Ren Nagasawa[b,d], Krishnan Mahalingam[b,1], Takamasa Hirai[b], Fuyuki Ando[b], Yuya Sakuraba[b,d], Satoru Kobayashi[a], and Ken-ichi Uchida[b,d,e,*]

[a.] *Faculty of Science and Engineering, Iwate University, Morioka 020-8551, Japan.*

[b.] *National Institute for Materials Science, Tsukuba 305-0047, Japan.*

[c.] *Center for Sustainable Materials and Interfacial Science, Iwate University, Morioka 020-8550, Japan*

[d.] *Graduate School of Science and Technology, University of Tsukuba, Tsukuba 305-8573, Japan.*

[e.] *Department of Advanced Materials Science, Graduate School of Frontier Sciences, The University of Tokyo, Kashiwa 277-8561, Japan.*

*Corresponding authors.

Email addresses: k.0yanagi444@gmail.com (K. Oyanagi), H.SEPEHRIAMIN@nims.go.jp (H. Sepehri-Amin), UCHIDA.Kenichi@nims.go.jp (K. Uchida).

[1] Prsent address: Sreenidhi University, Hyderabad, 501301, India.

Thermoelectric conversion based on the anomalous Nernst effect (ANE) is promising for energy harvesting as its transverse geometry enables the design of large-scale thermoelectric devices with simple structures. While topological ferromagnets, typically single crystals, exhibit large ANE, achieving high conversion performance remains challenging because it also requires high electric conductivity and low thermal conductivity. Here, we report enhanced transverse thermoelectric conversion performance in polycrystalline topological ferromagnet $Co_2MnGa$ (CMG) prepared by spark plasma sintering. Optimization of the sintering conditions for CMG leads to the anomalous Nernst coefficient of 7.5 μV $K^{-1}$ at room temperature, comparable to the highest value reported for the single crystals, and simultaneously reduces its thermal conductivity by 34% compared to that of the single crystals without affecting the electric conductivity. Owing to the transport properties that overcome conventional trade-off relations, our optimized CMG slab shows the record-high value of the dimensionless figure of merit for ANE at room temperature. Detailed nano/microstructure characterizations and first-principles phonon calculations clarify the unconventional dependence of the transport properties on the degree of crystalline ordering and morphology of crystal-domain boundaries. The results reveal the potential of polycrystalline topological materials for transverse thermoelectric applications and suggest alternative strategies to nanostructuring for enhancing thermoelectric performance.

## 1. Introduction

The anomalous Nernst effect (ANE) refers to the conversion of a longitudinal heat current into a transverse electric field in a magnetic material with spontaneous magnetization. Because of the transverse geometry of the thermoelectric conversion, ANE enables the construction of thermoelectric devices with convenient scalability and easy fabrication [1-4]. The output voltage of ANE increases simply by increasing a length of the device perpendicular to the applied temperature gradient, realizing a large-scale and junction-less thermoelectric module, which cannot be achieved by the Seebeck effect used in conventional thermoelectric modules. However, low thermoelectric conversion performance of ANE prohibits practical applications for thermoelectric devices.

Heusler alloys have been studied for several applications including spintronic devices with high Curie temperatures and tunable half-metallicity [5], the conventional thermoelectric devices with large Seebeck coefficients [6], and solid-state refrigeration with large magnetocaloric effects and magneto shape memory alloys with giant magnetic-field-induced strain effects [7]. Recently, magnetic Heusler alloys have been getting much attention for the transverse thermoelectric applications requiring lager ANE. For instance, a ferromagnetic Weyl semimetal $Co_2MnGa$ (CMG) with the $L2_1$ fully ordered Heusler structure, one of topological magnets exhibiting large anomalous Nernst coefficient $S_{ANE}$ due to non-trivial band structures [8-36], shows the largest $S_{ANE}$ value around room temperature [13,14,18,19,23]. However, ANE in CMG has been measured mostly in single crystals [13,18,19] ($S_{ANE}$ = 6.0 ~7.9 μV $K^{-1}$) and epitaxial thin films [14,23] ($S_{ANE}$ ~ 6.2 μV $K^{-1}$), which are not compatible to scalability and mass production.

For applications, polycrystalline slabs with large $S_{ANE}$ are promising to construct low-cost and large-scale devices based on ANE [37-42]. Owing to small anisotropy in the transport properties along the crystal orientation [13] and robustness to the grain boundary scattering [38], giant ANE has been reported not only in the polycrystalline films [29,34] but also in the polycrystalline slabs made by using a spark plasma sintering (SPS) method [37,38,41]. SPS is a versatile pressure-assisted sintering method in short time and low energy cost, and thus useful for controlling the microstructure in polycrystalline thermoelectric materials to improve their thermoelectric performance [43-45]. Furthermore, Ravi et al. demonstrated the remarkable change of $S_{ANE}$ in Fe-based alloys by microstructure engineering [46]. Therefore, the comprehensive understanding of the transport properties and detailed microstructure is crucial for the improvement of the performance of ANE in polycrystal materials for thermoelectric applications based on ANE.

In this study, we prepared polycrystalline CMG slabs with various sintering conditions using the SPS method and investigated their transport properties to optimize thermoelectric conversion performance of ANE, which is often evaluated by the dimensionless figure of merit [47-49]:

$$z_{\mathrm{ANE}}T = \frac{S_{\mathrm{ANE}}^2\sigma}{\kappa}T, \quad (1)$$

where $\sigma$ is the electric conductivity and $\kappa$ the thermal conductivity at absolute temperature $T$. The CMG slab sintered at a high temperature and high pressure exhibits $S_{\mathrm{ANE}}$ ~ 7.5 μV $\mathrm{K}^{-1}$, which is comparable to the best value in the single- and poly-crystalline CMG slabs at room temperature. We also achieved the decrease of $\kappa$ with maintaining sufficiently large $\sigma$ and $S_{\mathrm{ANE}}$, resulting in a record-high value of $z_{\mathrm{ANE}}T$ around room temperature among all the bulk magnetic materials reported so far. We then performed detailed nano/microstructure analysis on the samples to clarify the influence of the nano/microstructure on the transport properties. Transmission electron microscopy (TEM) results revealed the importance of the crystal growth of the $L2_1$ ordered phase within the samples for showing large $S_{\mathrm{ANE}}$ even if its existence cannot be detected by the X-ray diffraction (XRD). In addition to the conventional $L2_1$ and $B2$ phases in CMG, we found an unknown crystal phase with a size of ten nanometers. This length scale coincides with the scale of mean free paths for phonons which mainly contribute to the thermal conductivity obtained by a first-principles calculation, suggesting that the nanometer-scale crystal boundaries increase the phonon scattering and cause small $\kappa$. Our results provide an insight into the correlation between the transport properties and nano/microstructures within the sample, which is important for the simultaneous optimization of ANE, electric conductivity, and thermal conductivity.

## 2. Experimental

The polycrystalline CMG slabs were prepared under various sintering conditions from powder of CMG ingots. The detailed recipe for the preparation of CMG ingots is described in Supplementary Information and Ref. 37. We crushed the CMG ingots using a mortar and planetary ball mill, followed by sieving the ball-milled CMG powder through a 63 μm mesh. An inductively coupled plasma optical emission spectrometer determined the composition of the CMG powder to be $\mathrm{Co_{52.1}Mn_{23.1}Ga_{24.8}}$. We loaded the CMG powder into a graphite die with a diameter of 10 mm and sintered it under various conditions. We examined nine sintering conditions combining three sintering temperatures of $T_{\mathrm{sinter}}$ = 600, 700, and 800°C with three maximum sintering pressures of $p_{\mathrm{max}}$ = 30, 60, and 90 MPa. We kept $T_{\mathrm{sinter}}$ and $p_{\mathrm{max}}$ for 10 min in all the conditions. In the following, we define the sample name as CMG($T_{\mathrm{sinter}}$, $p_{\mathrm{max}}$) for each sintering condition. The sintered CMG slabs were cut into a rectangular shape with a length of 9-10 mm, width of 2 mm, and thickness of 0.4-0.9 mm for measuring the transport properties and magnetization $M$, and into a disc with a diameter of 10 mm and thickness of 1 mm for measuring the thermal diffusivity, using a diamond wire saw. Note that the variation in the shape of the rectangular samples does not affect the measurement results.

We measured ANE at room temperature and atmospheric pressure using a homemade sample holder (Fig. 1a). The sample holder consists of two anodized Al plates. The large one works as a thermal bath. The small one is equipped with a chip heater and is attached to the large one but thermally isolated from it by inserting a Bakelite board to create a temperature difference between the two plates. A sample was bridged between the plates and fixed with a high-thermal conductivity adhesive sheet. A uniform temperature difference along the $x$ direction was generated using the chip heater. Because of the thermal resistance between the sample and plates due to the presence of adhesive sheets, the temperature differences between the ends of the sample and between the plates are different [50]. To accurately determine $S_{\mathrm{ANE}}$, we estimated the actual value of the temperature gradient in the $x$ direction $\nabla_x T$ directly from a temperature-profile image at the surface of the sample coated with black ink using an infrared camera. We measured the thermoelectric voltage along the $y$ direction $V_y$, i.e., the sample width direction, with the magnetic field **H** along the out-of-plane direction and converted $V_y$ into the transverse electric field $E_y = V_y/w = S_{\mathrm{ANE}}\nabla_x T$, where $w$ is the sample width. The anomalous Hall effect was measured in the same setup for the ANE measurement, where a charge current was applied instead of the thermal gradient. The Seebeck coefficient and longitudinal electric conductivity were measured using a Seebeck Coefficient/Electric Resistance Measurement System (ZEM-3, Advance Riko, Inc.), which is similar to the system used in Ref. 51. The sample is clamped by two metallic blocks and is attached by two R-type (PtRh-Pt) thermocouple probes with the distance of 6 mm in a furnace under a He atmosphere. To measure the Seebeck coefficient, a temperature gradient is applied to the sample by heating one metallic block, and the resultant temperature difference and thermoelectric voltage at the same positions were measured simultaneously using the probes. By applying a charge current to the sample through the metallic blocks and measuring the voltage between the probes, we performed the standard direct current four-terminal method for measuring the longitudinal electric conductivity. The longitudinal resistivity of our samples is in the range from 1 μΩ m to 4 μΩ m (see Fig. S1a). The thermal conductivity was estimated by multiplying the thermal diffusivity, specific heat, and density. The disc shaped samples coated with black ink were used for the laser flash method (LFA1000, Linseis Messgeraete GmbH) to measure the thermal diffusivity.

The specific heat and density were measured using differential scanning calorimetry (DSCvesta, Rigaku Holdings Corp.) and Archimedes method, respectively. Note that the measurement results are not affected by the direction in which the transport properties were measured on the sample because of the homogeneity of the polycrystal slab [38]. The magnetization as a function of temperature was measured in the range from 300 K to 850 K using a superconducting quantum interference device equipped with a vibrating sample magnetometer (SQUID-VSM, Quantum Design) and from 5 K to 350 K using a superconducting quantum interference device (SQUID) magnetometer (MPMS-5L, Quantum Design), respectively. Scanning electron microscopy (SEM) was performed using a Carl Zeiss CrossBeam 1540EsB microscope equipped with an energy-dispersive spectroscopy (EDS) detector. Scanning transmission electron microscopy (STEM) was conducted using a Titan G2 80-200 (FEI) with a probe aberration corrector. The lift-out method was used to prepare the TEM specimens using a focused ion beam system Helios G4-UX DualBeam (FEI). The surfaces of the samples were observed using an ultra-low voltage SEM JSM-7800F Prime (Jeol Ltd.).

## 3. Results and discussion

Figure 1b shows the $H$ dependence of $E_y$ for CMG(800, 60) at different $\nabla_x T$ values. Clear $E_y$ signals appear by applying $\nabla_x T$ and $H$. The magnitude of the $E_y$ signal increases by increasing $\nabla_x T$ and its sign reverses with respect to the sign of $H$. By increasing $|H|$, the $E_y$ signal is saturated at around $|\mu_0 H| \sim 0.8$ T, which is consistent with the saturation field of $M$ (see the inset to Fig. 1b), indicating that ANE mainly generates the $E_y$ signal. In this setup, an $H$-linear signal due to the ordinary Nernst effect also appears. We subtracted the ordinary Nernst component by linear extrapolation for the results above the saturation field, represented as the colored dashed lines in Fig. 1b, and obtained the ANE component $E_{\mathrm{ANE}}$ at $H = 0$ (see the colored dots). We estimated $S_{\mathrm{ANE}} = E_{\mathrm{ANE}}/\nabla_x T$ to be ~7.5 μV K$^{-1}$ for CMG(800, 60) by a linear fit to the $\nabla_x T$ dependence of $E_{\mathrm{ANE}}$ in Fig. 1c. We obtained $S_{\mathrm{ANE}}$ for all the samples through the same procedure.

We show the $H$ dependence of $E_y$ for CMG(600, $p_{\mathrm{max}}$), CMG(700, $p_{\mathrm{max}}$), and CMG(800, $p_{\mathrm{max}}$) in Figs. 2a, 2b, and 2c, respectively. We found clear anomalous Nernst signals appear in all the samples and summarized their estimated $S_{\mathrm{ANE}}$ values in the left panel of Fig. 2d at room temperature. $S_{\mathrm{ANE}}$ monotonically increases by increasing $T_{\mathrm{sinter}}$, and the relationship of $S_{\mathrm{ANE}}$ at the same $T_{\mathrm{sinter}}$ value is CMG($T_{\mathrm{sinter}}$, 30) < CMG($T_{\mathrm{sinter}}$, 60) ~ CMG($T_{\mathrm{sinter}}$, 90). All the samples show relatively large $S_{\mathrm{ANE}}$ (> 2 μV K$^{-1}$), and the maximum $S_{\mathrm{ANE}}$ value of 7.5 μV K$^{-1}$ in CMG(800, 60) is the largest among all the values previously reported in polycrystalline magnets 29,30,32,34,36-38,46,48,49,52-54] including topological materials. More importantly, the maximum $S_{ANE}$ value in the polycrystalline CMG slab is comparable to or even larger than that in the single-crystalline CMG slabs [13,18,19] (see the blue and red dashed lines in the left panel of Fig. 2d).

To further investigate large ANE in the polycrystalline CMG slabs, we estimated the anomalous Nernst conductivity $\alpha_{xy}$. The left panel of Fig. 2e shows the obtained $\alpha_{xy}$ values in all the samples using the formula $S_{ANE} = \rho_{xx}\alpha_{xy} - (\rho_{AHE}S_{SE}/\rho_{xx})$ with the measured longitudinal resistance $\rho_{xx}$, anomalous Hall resistivity $\rho_{AHE}$, and Seebeck coefficient $S_{SE}$ (see Fig. S1). $\alpha_{xy}$ depends on the sintering condition and its trend is similar to that of $S_{ANE}$, while both $S_{ANE}$ and $\alpha_{xy}$ are independent of the saturation magnetization $M_s$ (see the right panels in Figs. 2d and 2e). The maximum value of ~ 3 A $m^{-1}$ $K^{-1}$ is comparable to that in the single-crystalline CMG samples [13,18,19]. This indicates the crucial role of the Berry curvature and electronic band structure at the Fermi level even in the polycrystals in showing large $S_{ANE}$, consistent with the recent findings [29,30,32-38,40,41].

We next focus on $z_{ANE}T$ to discuss the transverse thermoelectric performance of the polycrystalline CMG slabs. $\sigma$ and $\kappa$, which is obtained using the density, specific heat, and thermal diffusivity (see Fig. S2), for all the samples are summarized in Figs. 3a and 3b. We found that $\sigma$ and $\kappa$ of the samples prepared at $T_{sinter}$ = 700°C and 800°C are larger than those of the sample prepared at $T_{sinter}$ = 600°C. Although $\sigma$ of the samples prepared at $T_{sinter}$ = 700°C and 800°C is comparable to that of the single-crystalline CMG samples, $\kappa$ is maximally ~34% smaller than that of the single crystal [19]. This indicates that we successfully reduced $\kappa$ without decrease of $\sigma$ and $S_{ANE}$, which is important for improving $z_{ANE}T$ (see Eq. 1). Fig. 3c shows the sintering condition dependence of $z_{ANE}T$ for all the samples, which is estimated from the results in Figs. 2d, 3a, and 3b. The overall trend is determined primarily by $S_{ANE}$. The CMG(800, 60) sample shows the maximum value of ~ $8.0 \times 10^{-4}$, which is surprisingly greater than that for the single-crystalline CMG slab ($z_{ANE}T$ ~ $2.0 \times 10^{-4}$ in Ref. 13 and ~ $6.6 \times 10^{-4}$ in Ref. 19 shown as the red and blue dashed lines, respectively). Furthermore, this value is much larger than $z_{ANE}T$ in other magnetic materials [12,13,18,19,21,27,46,48,49,52-54] exhibiting large ANE, such as the single-crystalline $Fe_3Ga$ slab [21] and the polycrystalline $SmCo_5$-type permanent magnets [48], summarized in Fig. 3d.

Now, we consider the origin of the high-performance of ANE in the sintered CMG slabs. We found that the samples sintered at $T_{sinter}$ = 600°C and 700°C have a lower relative density, the ratio of the measured density to the theoretical density, than that of the samples sintered at 800°C (see Fig. S2) and a rough surface with remaining pores and microparticles (characterized by ultra-low voltage SEM as shown in Fig. S3). These indicate that the samples sintered at lower $T_{\mathrm{sinter}}$ are insufficiently densified, resulting in small $S_{\mathrm{ANE}}$ in CMG(600, $p_{\mathrm{max}}$) and CMG(700, 30). In fact, the CMG ingot annealed at 600°C with the relative density nearly of 100% exhibits $S_{\mathrm{ANE}}$ of 5.4 μV $K^{-1}$, which is much larger than that of the insufficiently densified CMG(600, $p_{\mathrm{max}}$) ($S_{\mathrm{ANE}} = 2 \sim 4$ μV $K^{-1}$), suggesting the correlation between $S_{\mathrm{ANE}}$ and the relative density, rather than the exposed temperature. On the other hand, although both CMG(800, 30) and CMG(800, 90) show the relative density of nearly 100% and dense morphology, CMG(800, 90) shows larger $S_{\mathrm{ANE}}$ than CMG(800, 30), indicating that the difference in the sample density cannot explain large ANE in CMG(800, 90). Therefore, to understand the origin of large $S_{\mathrm{ANE}}$ in CMG(800, 90), we performed the micro- and nano-scale structure analysis, i.e., SEM, energy dispersive X-ray spectrometry (EDS), and high-resolution high-angle annular dark field (HAADF) scanning transmission electron microscopy (STEM), on two samples.

Figures 4a and 4b show the SEM-EDS maps of Co, Mn, and Ga for the CMG(800, 90) and CMG(800, 30) slabs. We found an inhomogeneous distribution of Mn in both samples where Mn segregation can be seen at the surface of particles close to the grain boundary region. In both samples, the typical grain size is comparable (a few to several tens of micrometers). This grain size is consistent with the size of the initial powder of CMG, which was sieved through a 63 μm mesh.

Figures 5a, 5b, and 5c show high resolution HAADF-STEM images and nano-beam electron diffraction patterns obtained from CMG(800, 90) and CMG(800, 30). Although the bulk XRD patterns show only fundamental diffraction peaks of the $A2$ phase (see Fig. S4), the nano-beam diffraction patterns in the i and ii regions in Fig. 5a indicate the presence of the $L2_1$ and $B2$ ordered phases by 111 and 002 superlattice reflections in the diffraction patterns obtained along [111] zone axis of matrix phase, respectively, indicating that the ordered structure of CMG(800, 90) varies in the nanoscale containing. We found that the electron beam diffraction pattern in the iii region is different from that of both $L2_1$ and $B2$ phases; an additional superlattice reflection appears diagonally on either side of the 002 spot indicated by the white allows in the bottom of the right panels of Fig. 5a. To closely see the crystal structure of this unconventional modulated phase, we show the magnified HAADF-STEM images for the $L2_1$, $B2$, and modulated phases in Fig. 5c. In comparison with the $L2_1$ and $B2$ structures, the modulated phase consists of the relatively displaced atoms in the diagonal direction (see red colored dots indicated by the white arrows in the right panel of Fig. 5c), similar to a martensite. Although the TEM observation gives the local information, CMG(800, 30) seems to have a larger amount of the modulated phase than CMG(800, 90) (see the lines of the displaced atoms

indicated by the white arrows in Fig. 5b), suggesting that the increase of the sintering pressure at high sintering temperature facilitates the transformation of the modulated phase into the $L2_1$ and/or $B2$ phases.

We also found clear difference between CMG(800, 30) and CMG(800, 90) in the $M$-$T$ curves shown in Fig. 6. A large thermal hysteresis appears in CMG(800, 30) within the $T$ range from 500 K to 750 K, while it almost disappears in CMG(800, 90). The observed thermal hysteresis is irrelevant to the magnetic ordering transition because $M$ appears at around 800 K. The onset temperature of $M$ for CMG(800, 90) is lower than CMG(800, 30), and CMG(800, 30) shows larger $M$ than CMG(800, 90). A similar thermal hysteresis has been observed in the thin film of CMG [55] and other Heusler alloys with the martensitic transformations caused by the distortion of the crystal structure [56-58]. In our case, the observed martensitic-transformation-like hysteresis can be caused by the diagonal displacement of the atoms in the modulated phase (see Fig. 5c). Although determination of the actual Curie temperature is difficult in our case due to the application of the large magnetic field, the onset temperature of $M$ at 1 T for CMG(800, 90) is closer to the literature Curie temperature for $L2_1$-type CMG [59] (~685 K) than that for CMG(800, 30). By measuring the $M$-$T$ curves at 1 T at low temperatures (see the inset to Fig. 6), we obtained $M_s$ at $T$ = 5 K for CMG(800, 30) as 4.2 $\mu_B$ f.u.$^{-1}$ and CMG(800, 90) as 4.1 $\mu_B$ f.u.$^{-1}$ CMG(800, 90) shows $M_s$ consistent with the experimental and theoretical values for $L2_1$-type CMG [59,60] (~4.1 $\mu_B$ f.u.$^{-1}$) and closer to the experimental value for the $B2$-type CMG film [61] at 4.2 K (~3 $\mu_B$ f.u.$^{-1}$) than that for CMG(800, 30). All the results indicate that the magnetic properties of CMG(800, 90) are more similar to those of $L2_1$/B2-type CMG than those of CMG(800, 30), suggesting the transformation from the modulated phase into the $L2_1$ and/or $B2$ phases. This interpretation is consistent with the TEM observation. Therefore, the sintering pressure at high sintering temperature affects the degree of the crystalline orders in the samples.

The above results confirm that the degree of the crystalline order is important for obtaining large $S_{ANE}$ [21,62,63]. We found that CMG(800, 90) exhibits the larger value of $S_{ANE}$ than that in CMG(800, 30), which has more modulated phase than the $L2_1$ and/or $B2$ phases. Because the theoretical origin of large ANE in CMG is the topological electronic band structure in the fully ordered $L2_1$ phase [13,18,62], our results suggest that ANE in the modulated phase seems to be small, and thus the reduction of the modulated phase is crucial for obtaining large $S_{ANE}$.

Now, we focus on the reduction of $\kappa$ in CMG(800, 30) and CMG(800, 90). The inset to Fig. 3b shows the sintering condition dependence of the nonelectronic thermal conductivity $\Delta\kappa = \kappa - \kappa_{el}$, where the electronic thermal conductivity $\kappa_{el}$ is estimated via the Wiedemann-Franz law with the free-electron Lorenz number of $2.44 \times 10^{-8}$ W $\Omega$ $K^{-2}$. Since the values of $\sigma$ in CMG(800, 30) and CMG(800, 90) are comparable to that in the single crystal [19] (see Fig. 3a), $\kappa_{el}$ does not contribute to the decrease of $\kappa$ in our samples. On the other hand, we found the sizable decrease in $\Delta\kappa$ compared with the single crystal's value (represented as the dotted lines). Phonon and magnon can contribute to $\Delta\kappa$ in magnetic materials [64]. However, we assumed the magnon contribution is negligibly small in CMG at room temperature because the experimental observation of magnon contribution has been typically at very low temperatures [65,66] and the magnetic damping constant of the polycrystalline CMG [67] is one order of magnitude larger than that of CoFe alloys which show measurable magnon contribution at room temperature [68]. Therefore, the thermal conductivity carried by phonons plays an important role in the decrease of $\kappa$.

A plausible mechanism of the decrease of $\Delta\kappa$ is the increase of phonon-boundary scattering caused by nano/microstructure [46-48]. To investigate the phonon thermal conductivity $\kappa_{ph}$, we carried out a first-principles calculation, whose details are described in Supplementary Information, with taking 3- and 4-phonon, and isotope scatterings into account and obtained cumulative $\kappa_{ph}$, which provides a useful insight into the reduction of $\kappa_{ph}$ by nano/microstructures. Fig. 7a shows cumulative $\kappa_{ph}$ as a function of the phonon mean free path $L$ for CMG at room temperature. Cumulative $\kappa_{ph}$ rapidly increases from $L \sim 10$ nm and is saturated to $\sim 23$ W $K^{-1}$ $m^{-1}$ above $L > 1$ μm. The saturation value is comparable to the value of $\kappa$ in the single crystal and much larger than that in our samples. Our calculation indicates that phonons with $L$ in the range from 10 nm to 100 nm mainly carry heat in CMG, and scattering centers with the size of such the scale can efficiently reduce $\kappa_{ph}$. However, this length scale is much smaller than the typical grain size with the order of 10 μm in both CMG(800, 30) and CMG(800, 90) shown in the SEM images in Fig. 7b (showing the SEM images for all the samples in Fig. S3). We ignored phonon-electron and phonon-magnon scatterings in the calculation. They increase the phonon scattering and play an important role for quantitative discussion on $\kappa$, but only make the length scale of the heat-carrying phonons shorter. Therefore, even when phonon-electron and phonon-magnon scatterings are taken into account, the grains of the order of 10 μm in size cannot be responsible for the decrease in $\Delta\kappa$ due to the phonon-boundary scatterings. On the other hand, recall that the CMG samples contain the crystal phase separation in the nanometer scale, as observed in Figs. 5a-c. The coincidence between $L$ of the heat-carrying phonons and the size of the crystal phase separation indicates that the crystal-domain boundary induces phonon scatterings, resulting in the decrease of $\Delta\kappa$. Our results suggest that phonon engineering using not only grain boundaries but also crystal-domain boundaries can increase the performance of thermoelectric materials.

## 4. Conclusion

In summary, we investigated ANE at room temperature in polycrystalline CMG slabs prepared by the SPS method in various sintering conditions. The maximum values of $S_{\mathrm{ANE}}$ and $\alpha_{xy}$ of our polycrystalline CMG slabs prepared at a high sintering temperature and pressure are comparable to those in the single-crystalline CMG slab and are largest at room temperature among polycrystalline magnetic materials. Furthermore, the optimized CMG slab achieved the record-high $z_{\mathrm{ANE}}T$ value of $8 \times 10^{-4}$ at room temperature, which is larger than that for the single-crystalline CMG samples, owing to the decrease of $\kappa$. The transport measurements and nano/microstructure analysis indicate that the degree of the crystalline order is important for obtaining large $S_{\mathrm{ANE}}$. Based on the calculation of the phonon transport spectrum and nanoscale structure analysis, we suggest the importance of the crystal-domain boundary for increasing phonon scatterings to decrease $\Delta\kappa$ and $\kappa$, which is a different strategy of the conventional phonon engineering using grain boundaries to increase phonon scatterings. Our results demonstrate that the integration of approaches used for development of the conventional thermoelectric materials [69], e.g., phonon engineering, nano-structuring, and fabricating bulk composite, will be also important to improve the performance of the magneto-thermoelectric devices.

**Acknowledgement**

The authors thank M. Isomura and K. Suzuki for technical supports, and S. J. Park for fruitful discussion. This work was supported by JST CREST "Creation of Innovative Core Technologies for Nano-enabled Thermal Management" (JPMJCR17I1), JST ERATO "Magnetic Thermal Management Materials" (JPMJER2201), JSPS KAKENHI Grant-in-Aid for Early-Career Scientists (21K14519), JSPS KAKENHI Grant-in-Aid for Scientific Research (S) (22H04965), NEC Corporation, and NIMS Joint Research Hub Program. The computations in the present work were performed using the Numerical Materials Simulator at NIMS.

**Declaration of Competing Interests**

The authors declare that they have no known competing financial interests or personal relationships that could have appeared to influence the work reported in this paper.

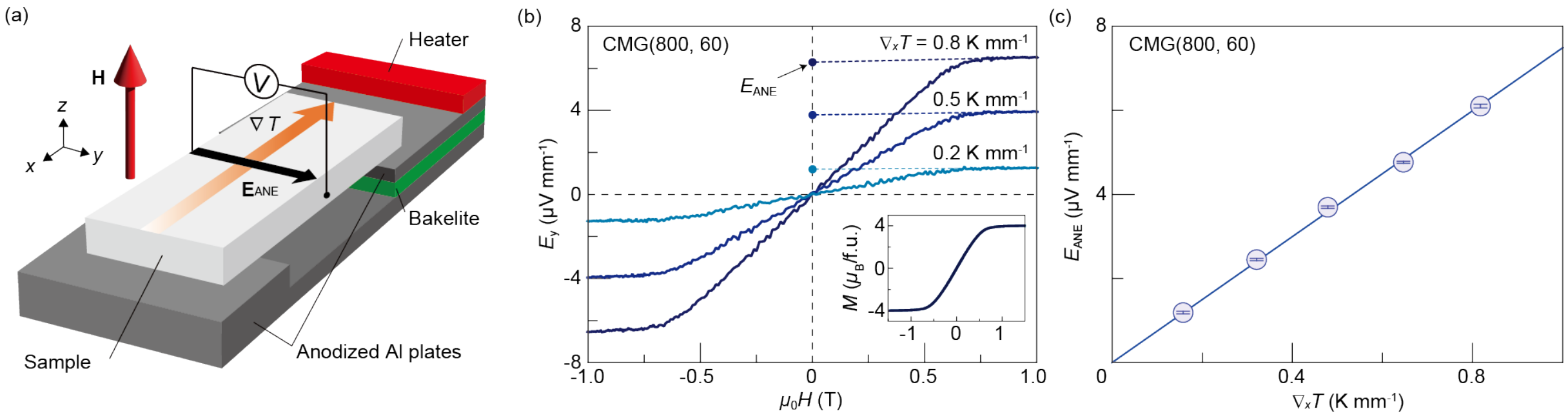

**Fig. 1** (a) Schematic illustration of the measurement setup. (b) $H$ dependence of $E_y$ in the CMG slab sintered at $T_{\text{sinter}}$ = 800°C and $p_{\text{max}}$ = 60 MPa, i.e., CMG(800, 60), for various values of $\nabla_x T$. The inset to (b) shows the $M$-$H$ curve of CMG(800, 60). The colored dashed lines in (b) show the linear extrapolation with the data in the high $H$ region, where $M$ is saturated (see the $M$-$H$ curve), for estimating $E_{\text{ANE}}$ represented as the colored dots at $\mu_0 H = 0$. (c) $\nabla_x T$ dependence of $E_{\text{ANE}}$. The solid line in (c) shows the result of a linear fitting, of which the slope corresponds to $S_{\text{ANE}}$. The error bars represent the 68% confidence level (±s.d.).

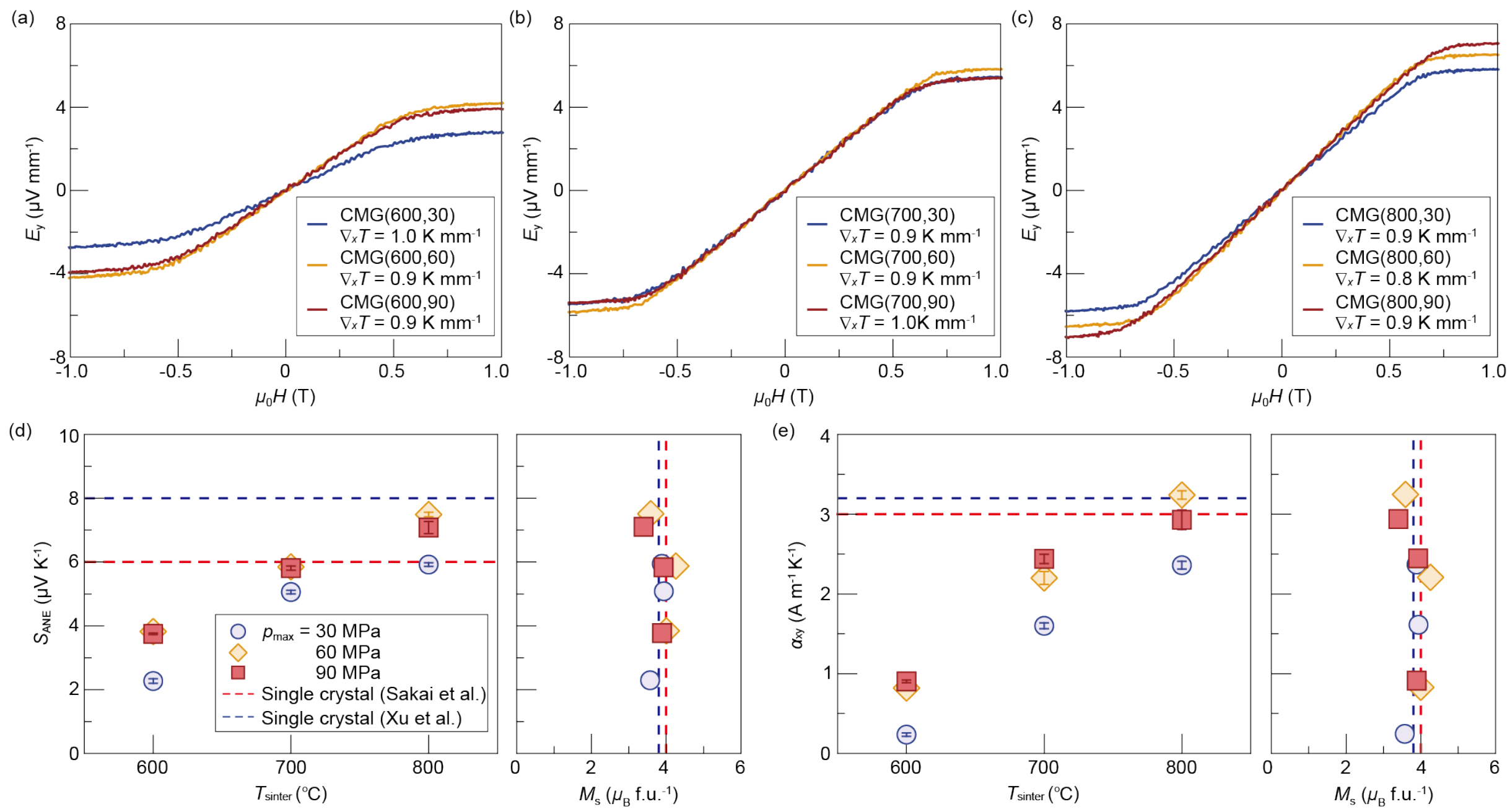

**Fig. 2** (a-c) $H$ dependence of $E_y$ in the CMG slab sintered at $T_{sinter}$ = 600°C (a), at $T_{sinter}$ = 700°C (b), and at $T_{sinter}$ = 800°C (c). (d) $T_{sinter}$, $p_{max}$, and $M_s$ dependences of $S_{ANE}$. (e) $T_{sinter}$, $p_{max}$, and $M_s$ dependences of $\alpha_{xy}$. The red and blue dashed lines in (d) and (e) correspond to the $S_{ANE}$, $\alpha_{xy}$, and $M_s$ values obtained in the single-crystalline samples in Refs. 13 and 19, respectively. The error bars represent the 68% confidence level (±s.d.).

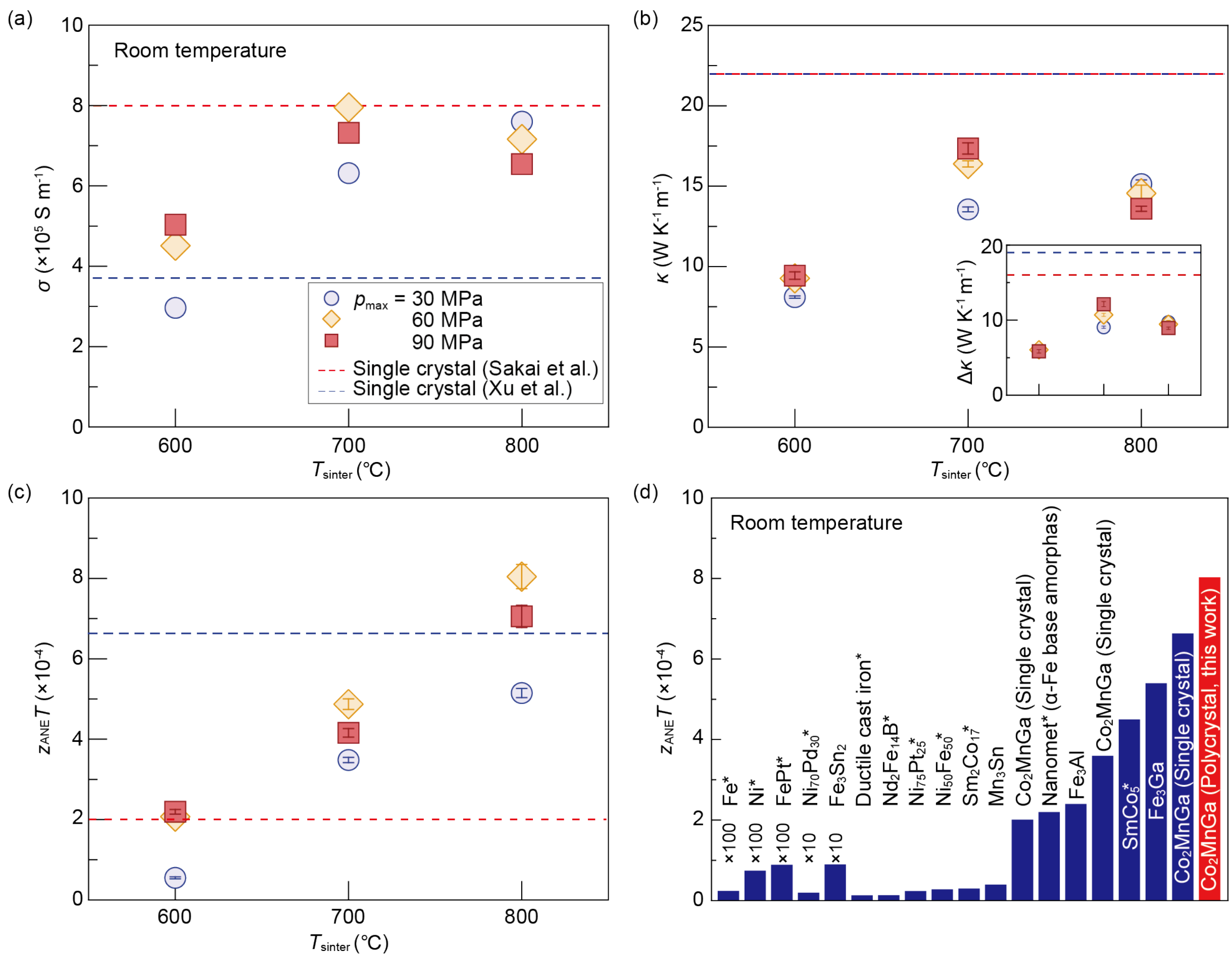

**Fig. 3** (a-c) $T_{\mathrm{sinter}}$ and $p_{\mathrm{max}}$ dependences of $\sigma$ (a), $\kappa$ (b), and $z_{\mathrm{ANE}}T$ (c) at room temperature ($T$ = 300 K). The inset to (b) shows the $T_{\mathrm{sinter}}$ and $p_{\mathrm{max}}$ dependences of $\Delta\kappa$. The red and blue dashed lines in (a-c) correspond to the $\sigma$, $\kappa$, $\Delta\kappa$, and $z_{\mathrm{ANE}}T$ values obtained in the single-crystalline samples in Refs. 13 and 19, respectively. The error bars represent the 68% confidence level (±s.d.). (d) Comparison of $z_{\mathrm{ANE}}T$ between our polycrystalline CMG slab (red bar) and the various bulk magnets (blue bars) around room temperature. $z_{\mathrm{ANE}}T$ for the materials with an asterisk are estimated using the experimental results of the anomalous Ettingshausen effect and the Onsager reciprocal relation.

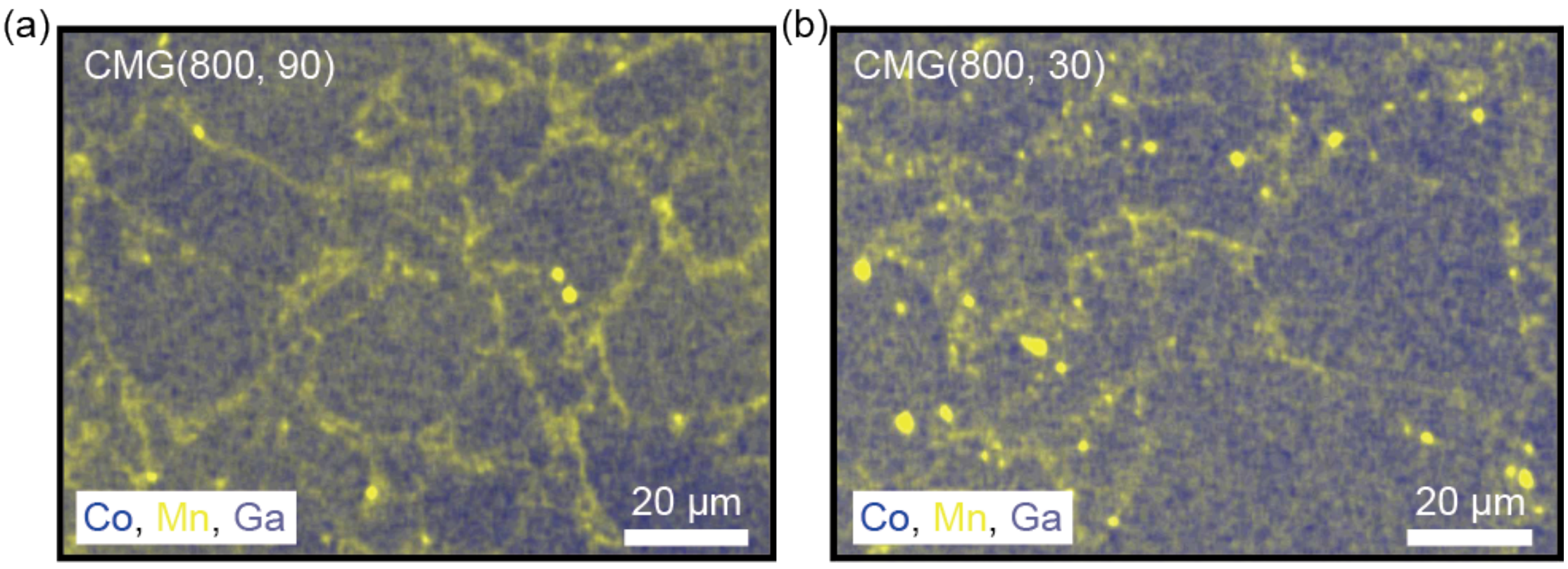

**Fig. 4** (a, b) SEM-EDS maps for CMG(800, 90) (a) and CMG(800, 30) (b).

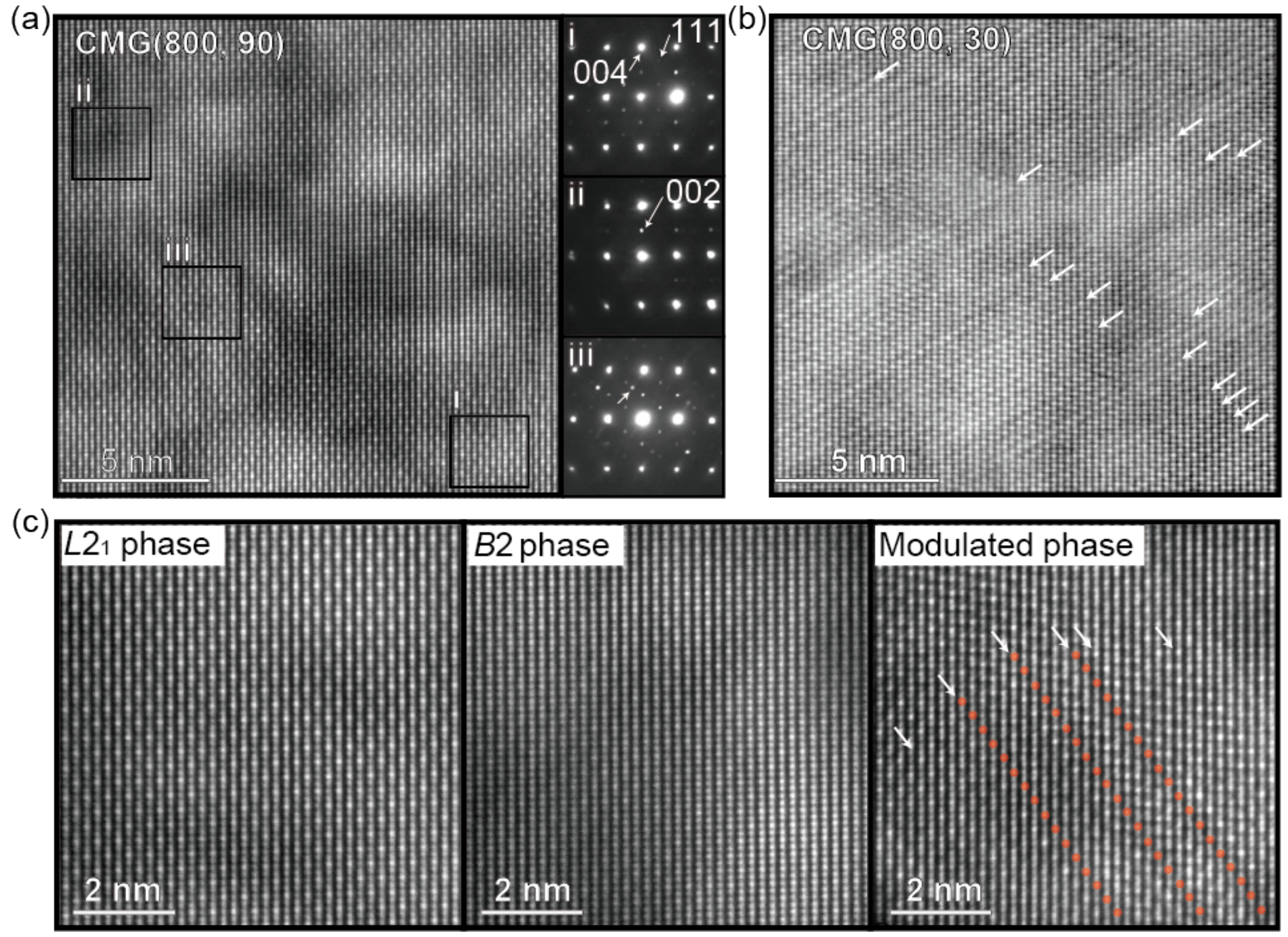

**Fig. 5** (a) High-resolution HAADF-STEM image obtained from CMG(800, 90) and electron-beam diffraction patterns obtained from the labelled regions. (b) High-resolution HAADF-STEM image from CMG(800, 30). (c) High-magnification HAADF-STEM images showing the $L2_1$, *B*2, and modulated phases.

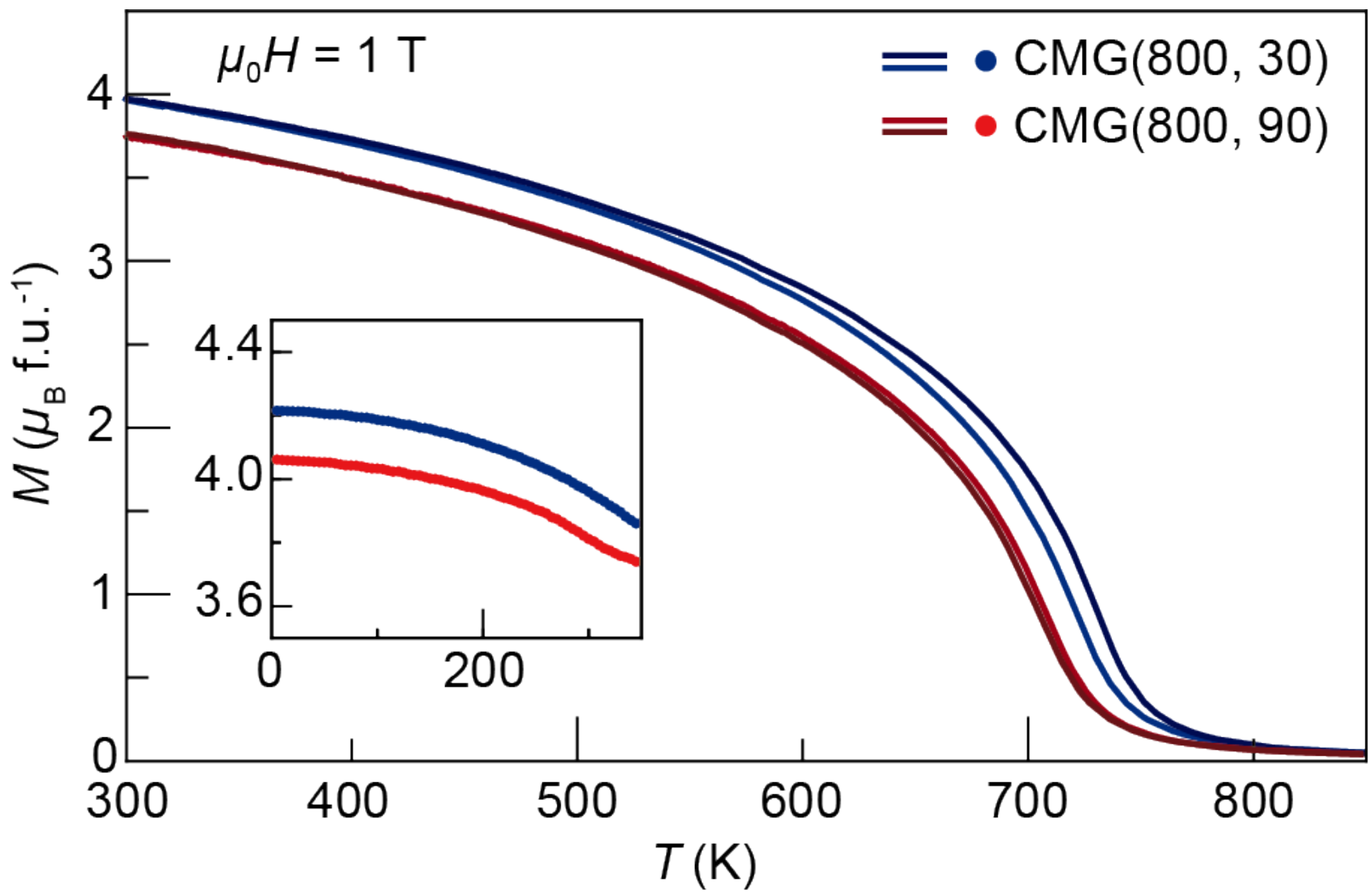

**Fig. 6** $T$ dependence of $M$ for CMG(800, 30) and CMG(800, 90) at $\mu_0H$ = 1 T. The inset shows the averaged values of $M$ in the $T$ range from 5 K to 350 K.

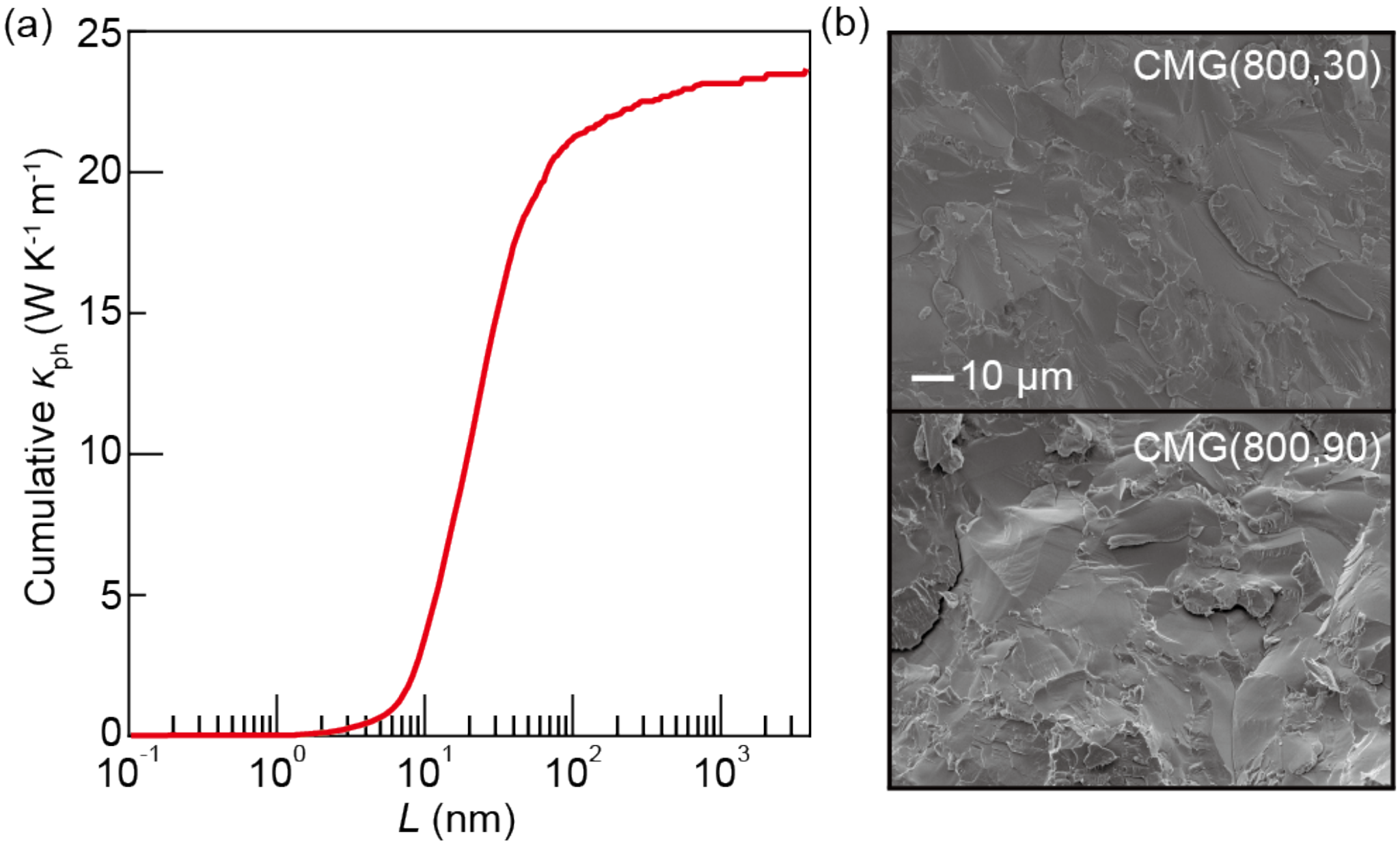

**Fig. 7** (a) Calculation of cumulative $\kappa_{ph}$ for CMG as a function of $L$ at $T$ = 300 K. (b) SEM images of the fracture surface of CMG(800, 30) and CMG(800, 90).